\documentclass[a4paper,11pt]{article}

\usepackage{jcappub}

\usepackage{tabularx}

\usepackage{mathtools}
\usepackage{tensor}
\usepackage{derivative}
\usepackage{bm}

\usepackage{soul}
\usepackage[normalem]{ulem}

\usepackage{suffix}

\usepackage[caption=false]{subfig}

\usepackage[dvipsnames]{xcolor}
\definecolor{myred}{rgb}{0.9,0.17,0.31}

\hypersetup{
    colorlinks=true,
    hyperfootnotes=true,
    linkcolor=myred,
    citecolor=cyan,
    urlcolor=magenta
}

\begin{document}

\title{Mass-Varying Neutrinos from an Inverse Symmetron}

\author[a]{Mainak Baidya,}
\author[b]{Øyvind Christiansen,}
\author[c]{Vitor da Fonseca,}
\author[d]{Eric V. Linder}
\author[e]{and David F. Mota}

\affiliation[a]{Department of Physics, Oklahoma State University,
Stillwater, OK 74078, USA}

\affiliation[b]{CEICO, Institute of Physics of the Czech Academy of Sciences,\\
Na Slovance 1999/2, 182 00 Prague 8, Czechia}

\affiliation[c]{Instituto de Astrof\'isica e Ci\^encias do Espa\c{c}o, Universidade de Lisboa,\\
Faculdade de Ci\^encias, Campo Grande, PT1749-016 
Lisboa, Portugal.}

\affiliation[d]{Berkeley Center for Cosmological Physics \& Berkeley Lab,
University of California,\\ Berkeley, CA 94720, USA}

\affiliation[e]{Institute of Theoretical Astrophysics, University of Oslo,\\
P.O. Box 1029 Blindern, N-0315 Oslo, Norway}

\emailAdd{mbaidya@okstate.edu}
\emailAdd{oyvinch@fzu.cz}
\emailAdd{vitor.dafonseca@alunos.fc.ul.pt}
\emailAdd{evlinder@lbl.gov}
\emailAdd{d.f.mota@astro.uio.no}

\abstract{
Neutrinos enter cosmology in different ways and are constrained by distinct observational probes across different epochs: as a relativistic species at high redshift, as a massive but clustering-suppressing component at low redshift, and as a particle physics observable in laboratory experiments. Low (verging on negative) bounds on neutrino mass from galaxy surveys motivate exploration of models where neutrinos may couple to dark energy, causing their mass to vary over cosmic evolution. If 
 the coupling 
 involves an inverse phase transition (symmetry broken, rather than restored, as neutrinos become nonrelativistic) this can tame instabilities in neutrino growth, appear as a lower neutrino mass in galaxy surveys, and add extra suppression to the matter power spectrum. We find that the late-time decoupling shuts down the fifth force and inhibits the excessive growth of neutrino perturbations, thereby eliminating linear-regime instabilities. The model may potentially address the Hubble tension via an early dark energy component localized around the time of recombination.
}


\maketitle
\flushbottom

\section{Introduction}
\label{sec:intro}

Neutrinos are elusive particles that pose challenges for both particle physics and cosmology. The Standard Model of particle physics originally assumed neutrinos to be massless, but terrestrial flavor oscillation experiments have demonstrated that they do possess mass \cite{Super-Kamiokande:1998kpq,T2K:2013ppw}. These measurements impose a lower bound on the sum of neutrino masses, ${\sum m_\nu>0.06\,\mathrm{eV}}$ (normal ordering) and ${\sum m_\nu>0.10\,\mathrm{eV}}$ (inverted ordering).

For its part, because neutrinos affect both the expansion history of the Universe and the growth of cosmic structure, cosmology provides an upper bound on the total neutrino mass by constraining the energy density of the cosmological neutrino background \cite{Lesgourgues:2006nd,10.3389/fphy.2017.00070}. This background is composed of thermal neutrinos that decoupled from the primordial plasma. Although the relic neutrinos have not yet been directly detected, they leave imprints at the background level and in the growth of large-scale structure. This allows cosmological observations to constrain their energy density, which translates into an upper limit on the sum of neutrino masses \cite{Workman:2022ynf}.

Recent baryon acoustic oscillations (BAO) measurements from the Dark Energy Spectroscopic Instrument (DESI), when combined with cosmic microwave background (CMB) data from the Planck satellite and the Atacama Cosmology Telescope, yield a tight upper limit of $\sum m_\nu<0.064\,\mathrm{eV}$ at 95\% confidence level (C.L.), assuming the $\Lambda\mathrm{CDM}$ model, which appears to disfavor the inverted mass ordering \cite{DESI:2024mwx,DESI:2025zgx}. When a broad range of late-time background probes are included, the DESI upper limit is further tightened to $\sum m_\nu<0.05\,\mathrm{eV}$ (95\% C.L.), below the minimum required by oscillation data, which reveals a $3\sigma$ tension between cosmology and terrestrial experiments \cite{Jiang:2024viw,Elbers:2025vlz}.

These results have significant implications for the $\Lambda$CDM paradigm because cosmological neutrino mass bounds are model-dependent. Expanding the framework of the concordance model with extra parameters (e.g., including additional relativistic species or altering the dark energy equation of state) can relax the cosmological constraints \cite{eBOSS:2020yzd,Valentino_2020}, potentially opening a window to new physics \cite{CosmoVerse:2025txj}. For instance, dynamical dark energy scenarios may alleviate the neutrino-mass tension. In the Chevallier-Polarski-Linder (CPL) parametrization \cite{param_a1,param_a2}, the corresponding $w_0w_a\mathrm{CDM}$ extension weakens the upper bound to $\sum m_\nu<0.16\,\mathrm{eV}$ \cite{DESI:2025zgx,Elbers:2025vlz}, or $\sum m_\nu<0.13\,\mathrm{eV}$ for non-phantom dark energy \cite{Vagnozzi:2018jhn}, restoring compatibility with oscillation experiments.

The neutrino tension may even hint at the presence of exotic cosmological scenarios, possibly connected to the physics of the dark sector. Among the proposed alternatives, mass-varying neutrino (MaVaN) models \cite{Gu:2003er,Fardon:2003eh,Peccei:2004sz,Brookfield:2005bz,Wetterich:2007kr, Amendola:2007yx,Ichiki:2007ng,Mota:2008nj,Geng:2015haa} are of particular interest because they provide a mechanism for relaxing cosmological neutrino mass constraints \cite{daFonseca:2023ury}. In these scenarios, neutrinos interact with a scalar field that may also drive the late-time acceleration of the Universe \cite{SupernovaSearchTeam:1998fmf,SupernovaCosmologyProject:1998vns}. Neutrino evolution proceeds as in the standard case until they become nonrelativistic. After this transition, the coupling to the scalar field induces a nontrivial evolution in both sectors,  causing the effective neutrino mass to vary with cosmic time, either increasing or decreasing in the recent Universe.

However, such models typically suffer from instabilities \cite{PhysRevD.72.065024, Bjaelde:2007ki, Franca:2009xp,Mandal:2019kkv}. When neutrinos acquire large enough masses, the associated fifth force can cause linear perturbations to grow rapidly, leading to neutrino clustering. In this work, we revisit the MaVaN framework by considering a scenario in which neutrinos are coupled to a symmetron scalar field. We show that the symmetron screening mechanism \cite{Hinterbichler:2010es, Hinterbichler:2011ca, Davis:2011pj, Silva:2013sla, Burrage:2018zuj, Hogas:2023vim, Christiansen:2023tfy, Vardanyan:2023jkm} still induces instabilities when a strong fifth force is activated following symmetry breaking. Motivated by this behavior, we propose an alternative scenario based on an inverse phase transition. In this setup, neutrinos decouple from the scalar field at late times, which suppresses the growth of perturbations and alleviates the instability problem at the background level.

In Section \ref{sec:model}, we introduce the neutrino-symmetron model, followed by a discussion of its background evolution. In Section \ref{sec:perturbations}, we analyze the model at the perturbation level, by integrating the linear fluctuation equations using the Boltzmann code CLASS~\cite{class2}. Section \ref{sec:inverse_PT} describes the inverse phase transition scenario and shows that it mitigates model instabilities. Finally, in Section \ref{sec:conclusions}, we summarize our results.

\section{Model description and background cosmology}
\label{sec:model}

We consider a spatially flat, homogeneous and isotropic Universe, whose expansion is parametrized by the scale factor $a$ associated with the Friedmann-Lema\^itre-Robertson-Walker (FLRW) metric, 
\begin{equation}
d s^2=a^2\left(-d \tau^2+d r^2+r^2 d \Omega^2\right)\,,
\end{equation}
where $\tau$ is the conformal time, that can also be expressed in terms of the cosmic time $t$ as, $d\tau=dt/a$. In natural units, $c=M_{\mathrm{Pl}}\equiv1/\sqrt{8\pi G_N}=1$, the Friedmann equations read 
\begin{equation}
\begin{aligned}
\mathcal{H}^2 & =\left(\frac{\dot{a}}{a}\right)^2=\frac{a^2}{3} \rho\,, \\
\dot{\mathcal{H}} & =-\frac{a^2}{6}(\rho+3 p)\,,
\end{aligned}
\end{equation}
where the dot denotes the derivative with respect to conformal time. $\mathcal{H}$ is the conformal Hubble parameter, and $\rho$ and $p$ represent the total energy density and pressure of the cosmic fluid, respectively. The cosmic fluid is composed of photons $(\gamma)$, massive neutrinos $(\nu)$, baryons $(b)$ and cold dark matter $(c)$, as well as a symmetron scalar field $(\phi)$ acting as a cosmological constant at late times when the field rolls to the potential minimum.

Our study focuses on a MaVaN model in which the mass of the neutrino evolves dynamically as a function of the symmetron $\phi$. Since, at leading order, cosmological observations primarily constrain only the sum of neutrino masses \cite{PhysRevD.73.123501,Font-Ribera:2013rwa}, we adopt for simplicity \cite{CORE:2016npo} a scenario with two massless neutrino species and one massive neutrino nonminimally coupled to the scalar field.

The effective neutrino mass is taken to be field-dependent and is parametrized as
\begin{equation}
\label{eq:m_nu}
m_\nu(\phi)=\hat{m}_{\nu}\,A(\phi)\,,
\end{equation} 
where $\hat{m}_\nu$ is the bare neutrino mass. For small field excursions, assuming $\mathbb{Z}_2$ symmetry, we can write 
the conformal factor 
\begin{equation}
\label{eq:conformal_factor}
    A(\phi)=1+\frac{\phi^2}{2M^2}+\mathcal{O}\left(\frac{\phi^4}{M^4}\right)\,,
\end{equation}
with $M$ a mass scale. 
We emphasize that this is not our final model, but a simplified case to illustrate key concepts. 
The bare neutrino mass \(\hat{m_{\nu}}\) then denotes the neutrino mass in the screened regime when $\phi\rightarrow0$. It also corresponds to the neutrino mass measured locally on Earth, assuming the symmetron is screened in the terrestrial environment. For the linear cosmological background, we define the present-day neutrino mass, $m_\nu^0$, as
\begin{equation}
    m_\nu^0\equiv m_\nu(z=0)\,.
\end{equation}

The neutrino energy density and pressure are given by \cite{Brookfield:2005bz}
\begin{align}
\label{eq:rho_nu}
    \rho_\nu&=\frac{1}{a^4}\int \frac{dq}{(2\pi)^3}d\Omega\,q^2\epsilon f_0(q)\,,\\
    p_\nu&=\frac{1}{3a^4}\int\frac{dq}{(2\pi)^3}d\Omega\,q^2f_0(q)\frac{q^2}\epsilon{}\,,
\end{align}
where
\begin{equation}
\label{eq:energy}
    \epsilon^2=q^2+m_\nu^2(\phi)a^2\,,
\end{equation}
is the neutrino comoving energy, and $f_0$ is the zero order Fermi-Dirac distribution function,
\begin{equation}
\label{eq:f_0}
f_0(q)=\frac{g_\nu}{e^{q / \Theta_\nu^0}+1}\,,
\end{equation}
which depends only on the comoving momentum $q$ because the neutrinos decouple from the thermal bath whilst they are still ultra-relativistic. Here $g_\nu$ stands for the number of neutrino degrees of freedom, and $\Theta_\nu^0$ is the present neutrino background temperature. 
Differentiating the equation of the neutrino density, Eq.~\eqref{eq:rho_nu}, with respect to conformal time leads to the neutrino continuity equation,
\begin{equation}
\dot{\rho}_\nu+3 \mathcal{H}\left(\rho_\nu+p_\nu\right)=\frac{d \ln m_\nu}{d \phi} \dot{\phi}\left(\rho_\nu-3 p_\nu\right)\,,
\end{equation}
which possesses a source term relating to the coupling between the neutrino and the symmetron field. In the above equation, $d\ln m_\nu=d\ln A=dA/A$, where
\begin{equation}
    dA = \frac{\phi\,d\phi}{M^2}\,.
\end{equation}
Assuming $dA\ll A\approx1$ 
we have $d\ln m_\nu\approx dA$, thus the continuity equation reads
\begin{align}
\label{eq:dev_A}
\dot{\rho}_\nu+3 \mathcal{H}\left(\rho_\nu+p_\nu\right)&=A_{,\phi}\,\dot{\phi}\left(\rho_\nu-3 p_\nu\right)\,, \\
&=\frac{\phi\,\dot{\phi}}{M^2}\left(\rho_\nu-3 p_\nu\right)\,.
\end{align}

As for the symmetron field, the self-interacting potential is given by \cite{Hinterbichler:2011ca}
\begin{equation}
\label{eq:potential}
V(\phi) = - \frac{1}{2} \mu^2 \phi^2 + \frac{1}{4} \lambda \phi^4+\frac{1}{4}\frac{\mu^4}{\lambda}+\rho_\Lambda
+\mathcal{O}(\phi^6)
\,,
\end{equation}
where $\mu$ is a mass scale that determines the symmetry-breaking threshold, $\lambda$ is a self-coupling parameter, and $\rho_\Lambda$ is the energy density of a cosmological constant. In the true vacuum, the potential minima have the cosmological constant energy.

The energy density and pressure associated with the canonical scalar field take the standard forms:
\begin{align}
\rho_\phi &= \frac{\dot{\phi}^2}{2a^2} + V(\phi)\,, \\
p_\phi &= \frac{\dot{\phi}^2}{2a^2} - V(\phi)\,.
\end{align}

The scalar field evolution is governed by the Klein-Gordon equation, modified to account for its coupling to neutrinos via the conformal factor $A(\phi)$, which enters through the scalar-dependent neutrino mass. The equation of motion reads\footnote{Note that our approach assumes $A>0$, $A\rightarrow 0$ gives a singular source unless treated differently.}:
\begin{align}
\label{eq:KG}
\ddot{\phi} + 2\mathcal{H} \dot{\phi} + a^2 V_{,\phi} &= -a^2 \frac{d \ln m_\nu}{d\phi} \left(\rho_\nu - 3p_\nu\right)\,, \\
&\approx -a^2 A_{,\phi} \left(\rho_\nu - 3p_\nu\right)\,, \\
\label{eq:KG_bis}
&\approx -a^2 \frac{\phi}{M^2} \left(\rho_\nu - 3p_\nu\right)\,.
\end{align}
Equivalently, the field continuity equation is
\begin{equation}
\label{eq:continuity_scalar}
    \dot{\rho}_\phi+3\mathcal{H}\left(\rho_\phi+p_\phi\right)=-\frac{\phi\dot{\phi}}{M^2}\left(\rho_\nu-3p_\nu\right)\,.
\end{equation}

The symmetron field evolves according to an effective potential given by
\begin{equation}
V_{\mathrm{eff}}(\phi)=\frac{1}{2}\left(-\frac{T_\nu}{M^2}-\mu^2\right)\phi^2+\frac{\lambda}{4}\phi^4+\frac{1}{4}\frac{\mu^4}{\lambda}+\rho_\Lambda\,,
\end{equation}
whose first derivative reads
\begin{align}\label{dVeff}
    V^\prime_{\mathrm{eff}}(\phi)=\left( -\frac{T_\nu}{M^2}-\mu^2 \right)\phi+\lambda\phi^3\,, 
\end{align}
where we use prime rather than ${}_{,\phi}$ to avoid double subscripts. The minima of the effective potential are located at,
\begin{equation}
\label{eq:evolving_minimum_standard}
    v_\pm = \pm v_0 \sqrt{\mathrm{max}\left\{1+\frac{T_\nu}{\rho_\star},
    \;0\right\}}\,,
\end{equation}
where we have defined the vacuum expectation value $v_0$ and the critical energy density $\rho_\star$ as,
\begin{equation}
\label{eq:v_0}
    v_0\equiv\frac{\mu}{\sqrt{\lambda}}\,,\quad\quad\rho_\star\equiv\mu^2M^2\,.
\end{equation}
We introduce two other useful phenomenological parameters, following Ref.~\cite{Hogas:2023vim},
\begin{align}
    \lambda_C &\equiv \frac{1}{m_\phi}\sim\frac{1}{\mu}\,,\\
    \label{eq:g}
    g &\equiv \frac{\mu\,M_{\rm Pl}}{\sqrt{\lambda}M^2}\,,
\end{align}
where $m_\phi$ is the field mass defined as $m_\phi^2=V''(\phi)$, $\lambda_C$ is the Compton wavelength giving the range of the fifth force mediated by the symmetron field, and $g$ is the dimensionless scalar-neutrino coupling ($M_\mathrm{Pl}=1$). The unscreened fifth-force strength relative to gravity is approximately
\begin{equation}
\frac{F_5}{F_{N}}
\simeq
2g^2\,,
\end{equation}
on scales shorter than the scalar Compton wavelength.

The stability of the $\phi = 0$ configuration is governed by the effective mass squared,
\begin{equation}
    \label{eq:meff_standard}
    m^2_{\mathrm{eff}} \equiv V''_{\mathrm{eff}}\big|_{\phi=0}
    = -\frac{T_\nu}{M^2} - \mu^2\,.
\end{equation}
The source term in the scalar field equation of motion is controlled by the trace of the neutrino stress-energy tensor, ${T_\nu=-\rho_\nu+3p_\nu}$, which vanishes for exactly massless or perfectly radiation-like neutrinos. For relativistic massive neutrinos, it is suppressed relative to the neutrino energy density, ${|T_\nu|\ll\rho_\nu}$, but it is not exactly zero. Its relevance for the scalar dynamics is determined by its size relative to the symmetry-breaking scale, ${\rho_\star=\mu^2M^2}$. Therefore, even a trace-suppressed neutrino component can affect the effective potential whenever ${|T_\nu|\gtrsim\rho_\star}$, and hence the dynamics of the field provided it is unfrozen.

\begin{figure*}[t]
    \centering
    \includegraphics[width=1\linewidth]{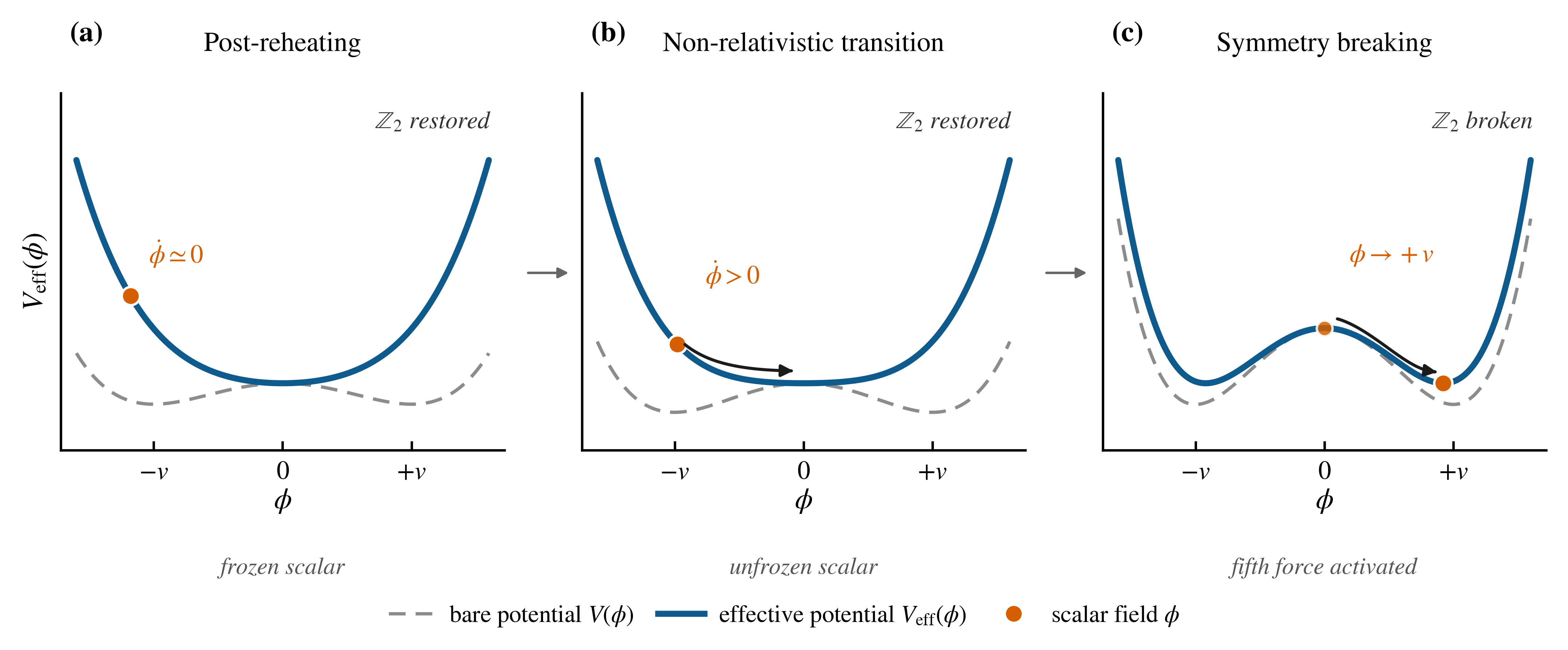}
    \caption{Behavior of the symmetron field coupled to neutrinos.}
    \label{fig:symmetron_behavior}
\end{figure*}

The behavior of the symmetron field is depicted in Fig.~\ref{fig:symmetron_behavior}. The homogeneous mode is initialized at ${\phi(z_{\mathrm{ini}})=-v_0}$, as seen in Panel (a). This can be realized in a scenario where the field perturbations are generated around the broken vacuum before reheating. Although $T_{\nu,\mathrm{ini}}\ll \rho_{\nu,{\mathrm{ini}}}$, the low phase transition energy $\rho_\star/\rho_{\nu,{\mathrm{ini}}}\ll (1-3\omega_{\nu,\mathrm{ini}})$, so that the $\mathbb{Z}_2$ symmetry is restored once sufficient neutrino density has been produced during reheating. The field does not necessarily track the corresponding minimum: if the scalar is overdamped by the expansion, its motion is suppressed by Hubble friction, until the field becomes dynamical as $ 2\mathcal{H}\dot\phi\ll \ddot\phi$, in Panel (b).

To note that if the field truly starts in the $\mathbb{Z}_2$ broken vacuum, we expect the Kibble-Zel'dovich mechanism to generate topological defects, in particular domain walls, with sufficiently low energy density as not to overclose the Universe. For simplicity, we initialize the perturbations adiabatically and neglect effects of topological defects.

As the Universe expands, \(\rho_\nu\) dilutes below the critical threshold, \({\rho_\nu<\rho_\star}\), thereby spontaneously breaking the $\mathbb{Z}_2$ symmetry, in panel (c). The field evolves toward one of two degenerate minima of the Mexican hat potential, $\pm v_0$. The redshift \(z_\star\) at which the phase transition occurs, and the trace-induced scalar-neutrino interaction becomes dynamically important, is given by the condition \({\rho_\nu(z_\star) = \rho_\star}\). Assuming approximately nonrelativistic neutrinos between ${z=0}$ and ${z=z_\star}$, the neutrino energy density evolves as
\begin{equation}
    \rho_\nu \propto a^{-3}A(\phi)\,.
\end{equation}
Since the conformal factor is close to unity,
\begin{align}
    1+z_\star
    &\simeq
    \left(
    \frac{\rho_\star}{\rho_\nu^0}
    \right)^{1/3}
    =
    \left(
    \frac{\mu^2M^2}{\rho_\nu^0}
    \right)^{1/3} \\
    &=
    \left(
    \frac{\mu^2M^2}{3H_0^2\Omega_\nu^0}
    \right)^{1/3}.
\end{align}
where \(\rho_\nu^0\) is the present-day neutrino energy density and $\Omega_\nu^0\equiv\rho_\nu^0/3H_0^2$ is the neutrino density parameter, which relates to the neutrino mass in the true vacuum as \cite{lesgourgues2013neutrino}
\begin{equation}
\label{eq:Omega_nu}
    \Omega_{\nu}^0 h^2=\frac{m_{\nu}^0}{93.14\,\mathrm{eV}}\,.
\end{equation}

\begin{figure*}[t]
    \centering
    \includegraphics[width=0.49\linewidth]{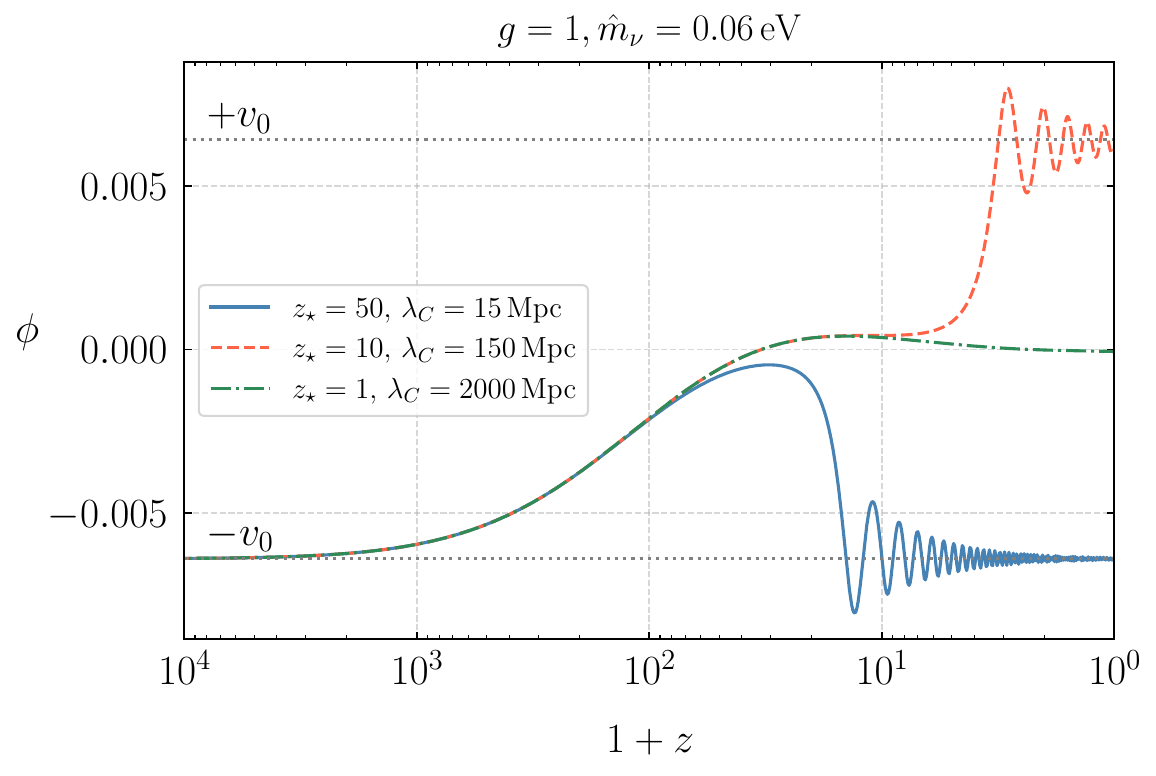}
    \hfill
    \includegraphics[width=0.49\linewidth]{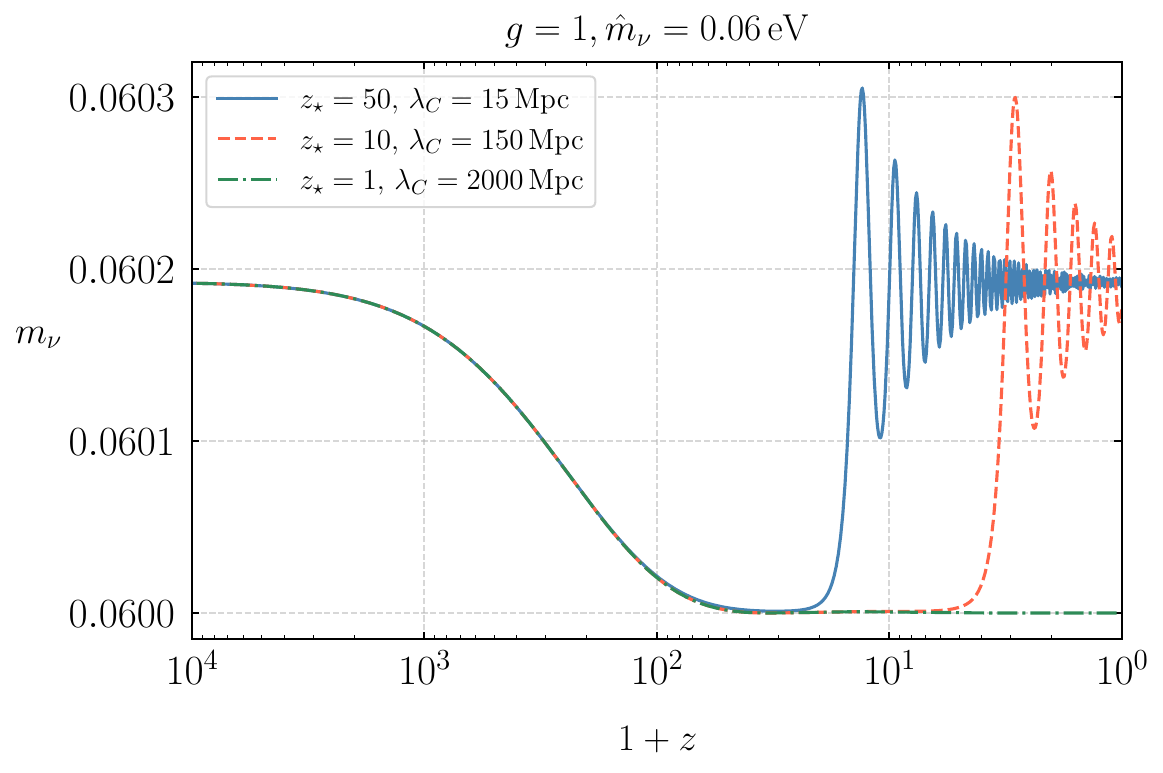}
    \caption{Left: Symmetron field evolution for three different symmetry-breaking redshifts: $z_\star=50, 10, 1$. The initial field value is fixed at $-v_0$, and $\lambda_C$ is adjusted in each case to yield similar minima $\pm v_0$ for a coupling constant $g = 1$. Right: Corresponding neutrino mass evolution.
    }
    \label{fig:phi_vs_z}
\end{figure*}

We have implemented the symmetron MaVaN model at the background level in the Einstein-Boltzmann code CLASS by modifying the comoving neutrino energy density in Eq.~\eqref{eq:energy}, as well as the scalar potential in Eq.~\eqref{eq:potential} and its equation of motion \eqref{eq:KG}. The left panel of Fig.~\ref{fig:phi_vs_z} shows a numerical example for $\hat{m}_\nu=0.06\,\mathrm{eV}$ confirming the redshift evolution of the scalar field for various redshifts of symmetry breaking.

As for the effective neutrino mass, its initial value follows from the Hubble frozen initial conditions,
\begin{equation}
\label{eq:neutrino_mass}m_\nu(z_\mathrm{ini})=\hat{m}_\nu\left(1+\frac{\mu^2}{2\lambda M^2}\right)\,,
\end{equation}
as depicted in the right panel of Fig.~\ref{fig:phi_vs_z}. Once Hubble friction reduces sufficiently, the field evolves toward zero in response to the trace-modified effective potential, and the mass decreases toward $\hat{m}_\nu$. If the symmetry-breaking redshift is set to ${z_\star=1}$, the field does not have time to exit the false vacuum and thus ${m_\nu^0\rightarrow\hat{m}_\nu}$. In contrast, for an earlier symmetry-breaking redshift, the field evolves toward one of the vacuum expectation values ${\pm v_0}$, and the mass of the coupled neutrino increases back to Eq.~\eqref{eq:neutrino_mass}. Therefore the present-day neutrino mass converges to
\begin{equation}
    m_\nu^0\rightarrow \hat{m}_\nu\left(1+\frac{\mu^2}{2\lambda M^2}\right)\,.
\end{equation}
The symmetron's self-interaction starts becoming dominant in the symmetron potential sooner for higher values of $z_\star$; for instance, the field leaves the symmetric phase earlier for ${z_\star=50}$ than for ${z_\star=10}$.

\section{Impact on perturbations}
\label{sec:perturbations}

In line with previous studies (e.g.\  \cite{Ichiki:2007ng,Brookfield:2005bz,Franca:2009xp}), the perturbations in the energy density and pressure of the interacting neutrinos are modified due to the coupling with the scalar field. They are given by:
\begin{align}
\label{eq:perturbed_neutrino_density}
\delta\rho_\nu &= \frac{1}{a^4} \int q^2dq\,d\Omega\,f_0 \left( \epsilon \Psi + \delta\phi A_{,\phi} \frac{m_\nu^2 a^2}{\epsilon} \right)\,, \\
\label{eq:perturbed_neutrino_pressure}
\delta p_\nu &= \frac{1}{3a^4} \int q^2dq \, d\Omega \, f_0 \frac{q^2}{\epsilon^2} \left( \epsilon \Psi - \delta\phi A_{,\phi} \frac{ m_\nu^2 a^2}{\epsilon} \right)\,,
\end{align}
where $A_{,\phi}=\phi/M^2$, $f_0$ is the unperturbed neutrino distribution function in Eq.~\eqref{eq:f_0}, $\Psi$ is the perturbation to the phase-space distribution, and $\delta\phi$ denotes the fluctuation of the symmetron field. The modifications arise from the scalar-dependent mass $m_\nu(\phi)$ through the conformal factor $A(\phi)$ defined in Eq.~\eqref{eq:m_nu}.

While the shear equation remains unchanged from the uncoupled case, the dipole moment equation for the neutrino hierarchy acquires an additional source term due to the coupling. In Fourier space, it reads
\begin{equation}
\label{eq:dipole}
\dot{\Psi}_1 = \frac{qk}{3\epsilon} \left(\Psi_0 - 2\Psi_2 \right) - \frac{qk}{3\epsilon} A_{,\phi} \, \delta\phi \frac{a^2 m_\nu^2}{q^2} \frac{d \ln f_0}{d \ln q}\,.
\end{equation}

We have extended the treatment of the non-cold dark matter (ncdm) component in the CLASS code \cite{Lesgourgues_2011} to incorporate the evolution of perturbations in the symmetron MaVaN model, by implementing Eqs.~\eqref{eq:perturbed_neutrino_density}, \eqref{eq:perturbed_neutrino_pressure} and \eqref{eq:dipole}.

We have also modified the perturbed Klein-Gordon equation in the code to account for the coupling between the symmetron field and neutrinos at the perturbation level,
\begin{align}
    \label{eq:perturbed_KG}
    \ddot{\delta\phi} &+ 2\mathcal{H}\dot{\delta\phi}
    + \left(k^2 + a^2 V_{,\phi\phi} \right)\delta\phi 
    + \tfrac{1}{2}\dot{h}\dot{\phi} 
    \notag =\\
    & -a^2 \big[
        A_{,\phi}(\delta\rho_\nu - 3\delta p_\nu)
        + A_{,\phi\phi}\,\delta\phi(\rho_\nu - 3p_\nu)
    \big]\,,
\end{align}
where $h$ denotes the trace of the spatial metric perturbation in synchronous gauge, following the conventions of Ref.~\cite{Ma:1995ey}, and $A_{,\phi\phi}=1/M^2$.

Solving for the evolution of linear density fluctuations with the Boltzmann code CLASS allows us to capture the impact of the symmetron-neutrino interaction at the level of cosmological perturbations. In particular, the coupling introduces an additional force acting on neutrinos once the symmetry is broken, thereby altering their clustering properties. For sufficiently large fifth-force strength and neutrino mass, the model exhibits instabilities similar to those reported in Ref.~\cite{PhysRevD.72.065024,Bjaelde:2007ki,Mota:2008nj,Franca:2009xp}.

\begin{figure}[t]
    \centering
\includegraphics[width=0.7\linewidth]{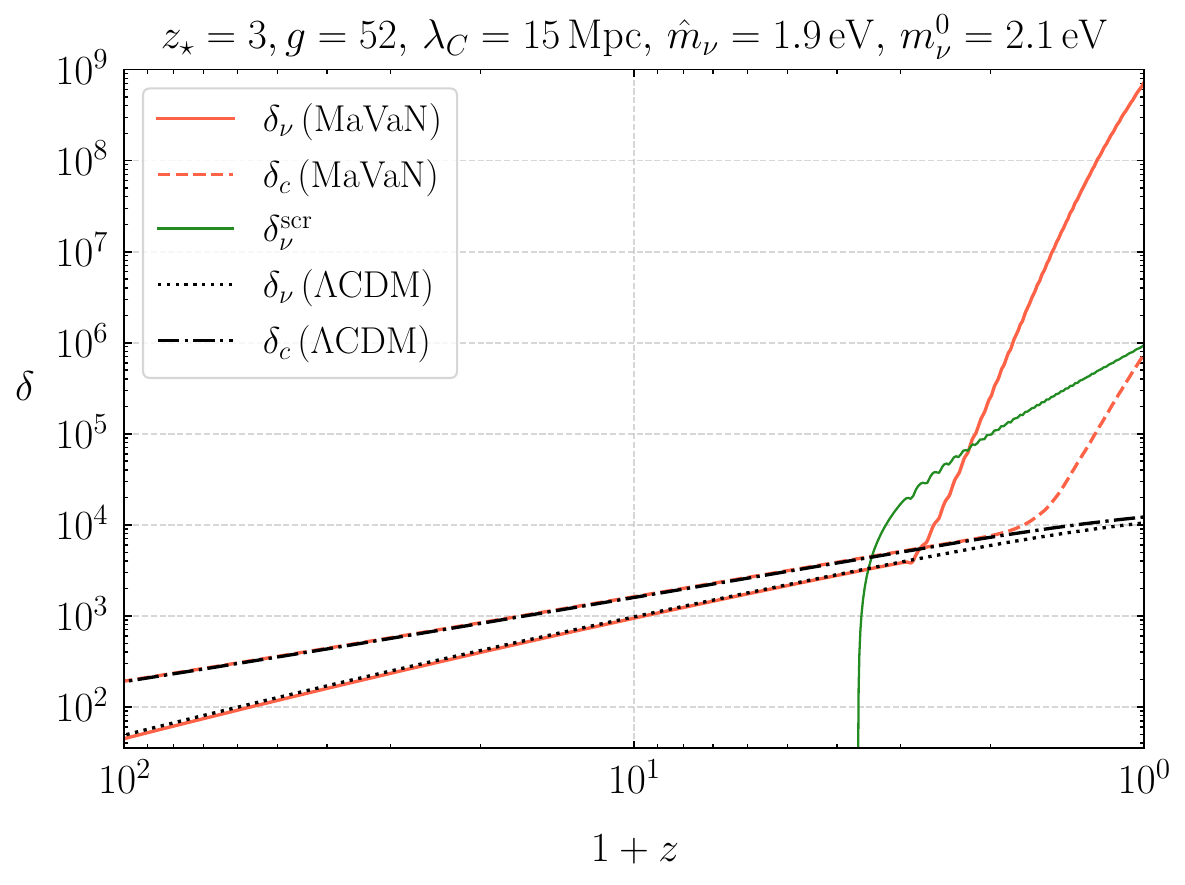}
    \caption{Evolution of the neutrino $(\delta_\nu)$ and cold dark matter $(\delta_c)$ density contrast for the MaVaN symmetron model (symmetry breaking around ${z_\star\sim3}$) and $\Lambda\mathrm{CDM}$ (with neutrinos of mass $m_\nu^0=2.1\,\mathrm{eV}$), at scale $k=0.1\,\mathrm{Mpc}^{-1}$, normalized to the primordial curvature amplitude \(\mathcal{R}=1\). The nonlinear screening threshold $\delta_\nu^{\rm{ scr}}$ in Eq.~\eqref{eq:screening_threshold} is indicated by the green curve.}
\label{fig:density_contrast}
\end{figure}

This is shown in Fig.~\ref{fig:density_contrast} at a scale $k=0.1\,\mathrm{Mpc^{-1}}$ and for the same parameter values that produce the instabilities in Ref.~\cite{Mota:2008nj}: $g=52$ and present-day neutrino mass $m_\nu^0=2.1\,\mathrm{eV}$. Similarly to that model the neutrino fluctuations grow rapidly around $z\sim3$, in the aftermath of symmetry breaking when the scalar-neutrino coupling becomes active. The effective mass of the scalar changes sign and the field settles into the symmetry-broken minimum, generating a non-vanishing coupling to neutrinos. Once the interaction turns on, neutrino perturbations experience an additional attractive force, which enhances their clustering beyond the standard gravitational contribution. The neutrino density contrast therefore increases sharply and can reach non-linear values within a short redshift interval. In this regime, neutrino overdensities begin to gravitationally source and drag dark matter perturbations.

However, the linear evolution studied here does not account for the non-linear screening mechanism of the symmetron. Sufficiently large neutrino overdensities can restore the symmetry locally, thereby suppressing the scalar-mediated force, once the local neutrino density exceeds the critical density of the phase transition. In the nonrelativistic regime, this condition can be approximated as a threshold on the neutrino density contrast,
\begin{equation}
\label{eq:screening_threshold}
\delta_\nu^{\rm scr}(z)
=
\frac{\rho_\star}{\bar\rho_\nu(z)}-1\,,
\end{equation}
where the overbar denotes the background energy density. For the parameter set considered here, this threshold is crossed after the onset of the linear instability, suggesting that non-linear screening effects should become relevant before the extremely large late-time fluctuations predicted by the purely linear calculation are reached. Whether the symmetron screening mechanism is strong enough to halt the instability and render this parameter regime observationally viable cannot be confirmed within the present linear treatment, as this question would require a dedicated non-linear analysis.

\begin{figure}[t]
    \centering
\includegraphics[width=0.7\linewidth]{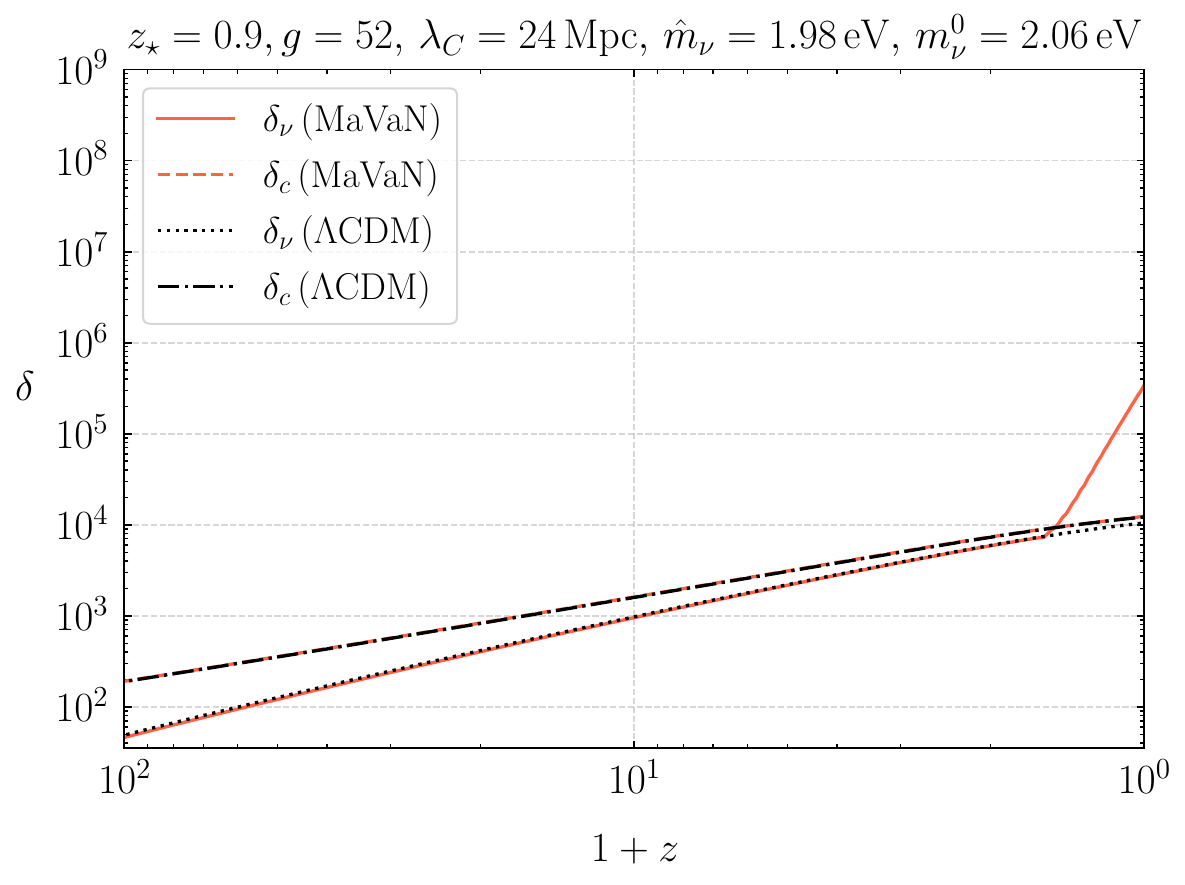}
    \caption{Evolution of the neutrino $(\delta_\nu)$ and cold dark matter $(\delta_c)$ density contrast for the MaVaN symmetron model (symmetry breaking around ${z_\star\sim0.9}$) and $\Lambda\mathrm{CDM}$ (with neutrinos of mass $m_\nu^0=2.1\,\mathrm{eV}$), at scale $k=0.1\,\mathrm{Mpc}^{-1}$, normalized to the primordial curvature amplitude \(\mathcal{R}=1\).}
\label{fig:density_contrast_delayed}
\end{figure}

Moreover, the onset of the instability depends on the time at which the phase transition takes place. Later transitions may postpone the instability so that its impact on the cold dark matter sector remains limited by the present epoch. This is illustrated in Fig.~\ref{fig:density_contrast_delayed}, where the symmetry breaking occurs only at ${z\sim0.9}$. In this case, the neutrino density contrast still undergoes a pronounced late-time enhancement while the cold dark matter overdensity does not exhibit runaway growth before $z=0$.

Since the model that is considered develops potentially problematic linear instabilities, we now explore an extension of this simple setup. A first possibility is to allow for a stronger variation of the effective neutrino mass.
In the minimal symmetron setup considered so far, the mass variation is controlled by a conformal factor \(A(\phi)\)  that was assumed to remain close to unity. 
For instance, in the example discussed above, the variation is only at the level of \(10\%\), with
\(\hat m_\nu = 1.9\,\mathrm{eV}\) and \(m_\nu^0 = 2.1\,\mathrm{eV}\).

While the conformal factor in Eq.~\eqref{eq:conformal_factor} was taken to be representative of general $\mathbb{Z}_2$-symmetric conformal couplings for small field excursions (and hence neutrino mass variation), a particularly motivated completion of it for larger field excursions is the exponential
\begin{equation}
\label{eq:conformal_factor_exponential}
A(\phi)=\exp\left({\frac{\phi^2}{2M^2}}\right)\,,
\end{equation}
so that the small-field limit is unchanged, while larger field excursions can generate a significantly stronger variation of the neutrino mass.

The evolution of the neutrino mass is changed with respect to the standard symmetron model, but the derivatives relating to the interaction that enter the equations are the same:
\begin{align}
    \frac{d\ln m_\nu}{d\phi}&=\frac{\phi}{M^2}\,,\\
    \frac{d^2\ln m_\nu}{d\phi^2}&=\frac{1}{M^2}\,.
\end{align}

Keeping the same coupling strength, \({g=52}\), and the same present-day neutrino mass,
\({m_\nu^0=2.1\,\mathrm{eV}}\), as in the previous case, the exponential conformal factor allows for a much larger mass variation. 
In our new setup, the bare neutrino mass is
\({\hat m_\nu = 3\times 10^{-7}\,\mathrm{eV}}\), which corresponds to the neutrino mass in the screened limit, \({\phi\to0}\). 
During the nonrelativistic regime before symmetry breaking, however, the background field does not reach exactly \({\phi=0}\). 
Consequently, given the actual evolution of \(\phi\), the minimum effective neutrino mass attained along the cosmological trajectory is instead ${m_{\nu,\min}\simeq 5\times 10^{-3}\,\mathrm{eV}}$.

\begin{figure}[t]
    \centering
    \includegraphics[width=0.7\linewidth]{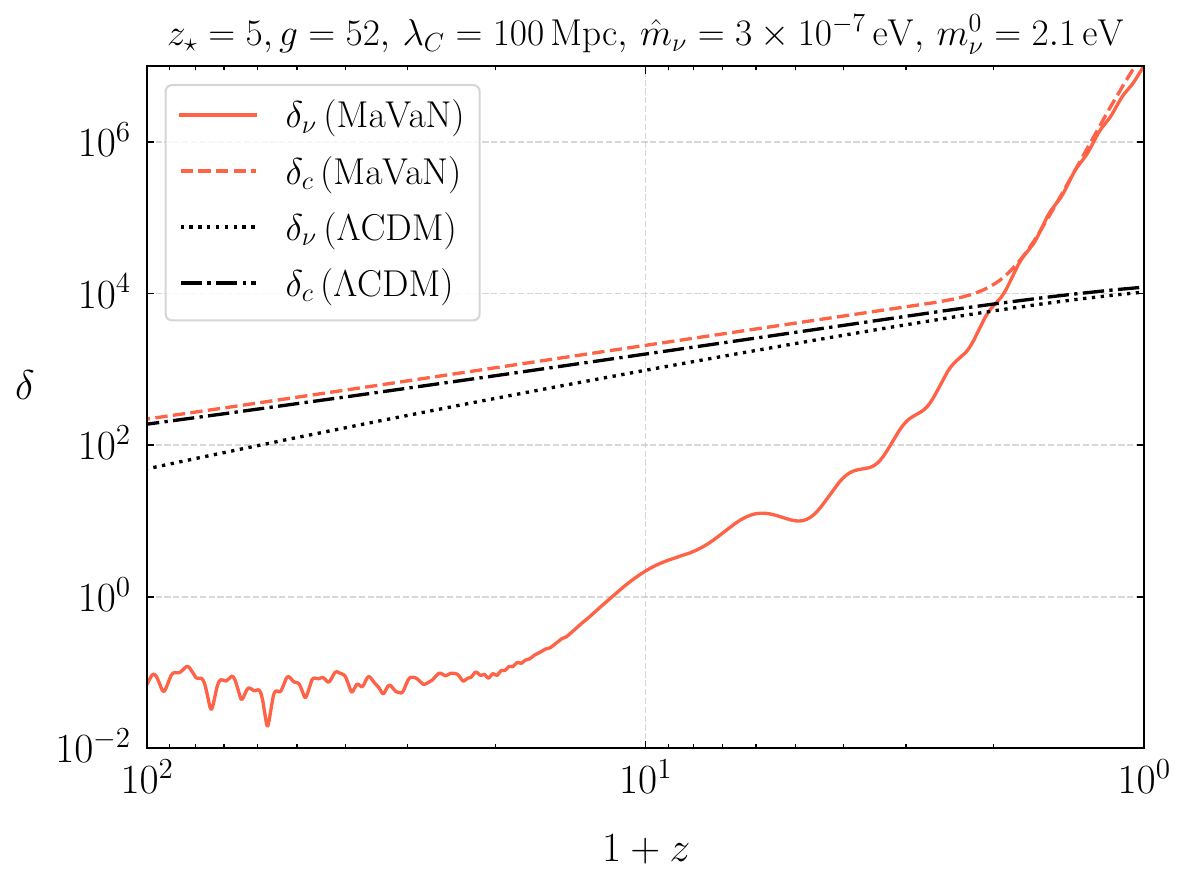}
    \caption{Evolution of the neutrino $(\delta_\nu)$ and cold dark matter $(\delta_c)$ density contrast for the MaVaN exponential conformal factor case, and for $\Lambda\mathrm{CDM}$ (with neutrinos of mass $m_\nu^0=2.1\,\mathrm{eV}$), at scale $k=0.1\,\mathrm{Mpc}^{-1}$, normalized to the primordial curvature amplitude \(\mathcal{R}=1\).}
    \label{fig:density_contrast_exponential}
\end{figure}

Despite this enhanced mass variation, the model with the exponential conformal factor still develops instabilities. 
In particular, the large late-time growth of neutrino and dark matter fluctuations persists, as shown in Fig.~\ref{fig:density_contrast_exponential}.
Note that before the onset of symmetry breaking, the effective neutrino mass is much smaller than in the corresponding \(\Lambda\mathrm{CDM}\) model with constant neutrino mass. 
This explains the enhanced growth of the cold dark matter density contrast seen in Fig.~\ref{fig:density_contrast_exponential} during that period.

One possible approach to ensuring model stability would be to phenomenologically construct an \textit{inverse phase transition} mechanism in which the scalar-neutrino coupling dynamically switches \textit{off} at late times. We will explore such a scenario in the next section. 

\section{Inverse phase transition model}
\label{sec:inverse_PT}

\subsection{Model description}

We want to construct a model where the fifth force is present at earlier times but naturally disappears at late time. To do so, we replace Eq.~\eqref{eq:conformal_factor_exponential} by a conformal factor with a negative exponential
\begin{equation}
\label{eq:conformal_factor_exponential_anti}
    A(\phi)=\exp\left({-\frac{\phi^2}{2M^2}}\right)\,,
\end{equation}
and replace the scalar potential in Eq.~\eqref{eq:potential} by
\begin{equation}
\label{eq:potential_anti}
V(\phi) = \frac{1}{2} \mu^2 \phi^2 + \frac{1}{4} \lambda \phi^4+\rho_\Lambda \,.
\end{equation}
With this choice, the effective neutrino mass becomes
\begin{equation}
\label{eq:mass_anti}
    m_\nu(\phi)=\hat m_\nu A(\phi)
    =\hat m_\nu \exp\left(-\frac{\phi^2}{2M^2}\right)\,.
\end{equation}
In the late-time restored phase, ${\phi\to0}$, so that ${A(\phi)\to1}$. Therefore, contrary to the previous models, the bare neutrino mass coincides with the present-day mass, ${\hat m_\nu = m_\nu^0}$, as will be shown below.

\begin{figure*}[t]
    \centering
    \includegraphics[width=1\linewidth]{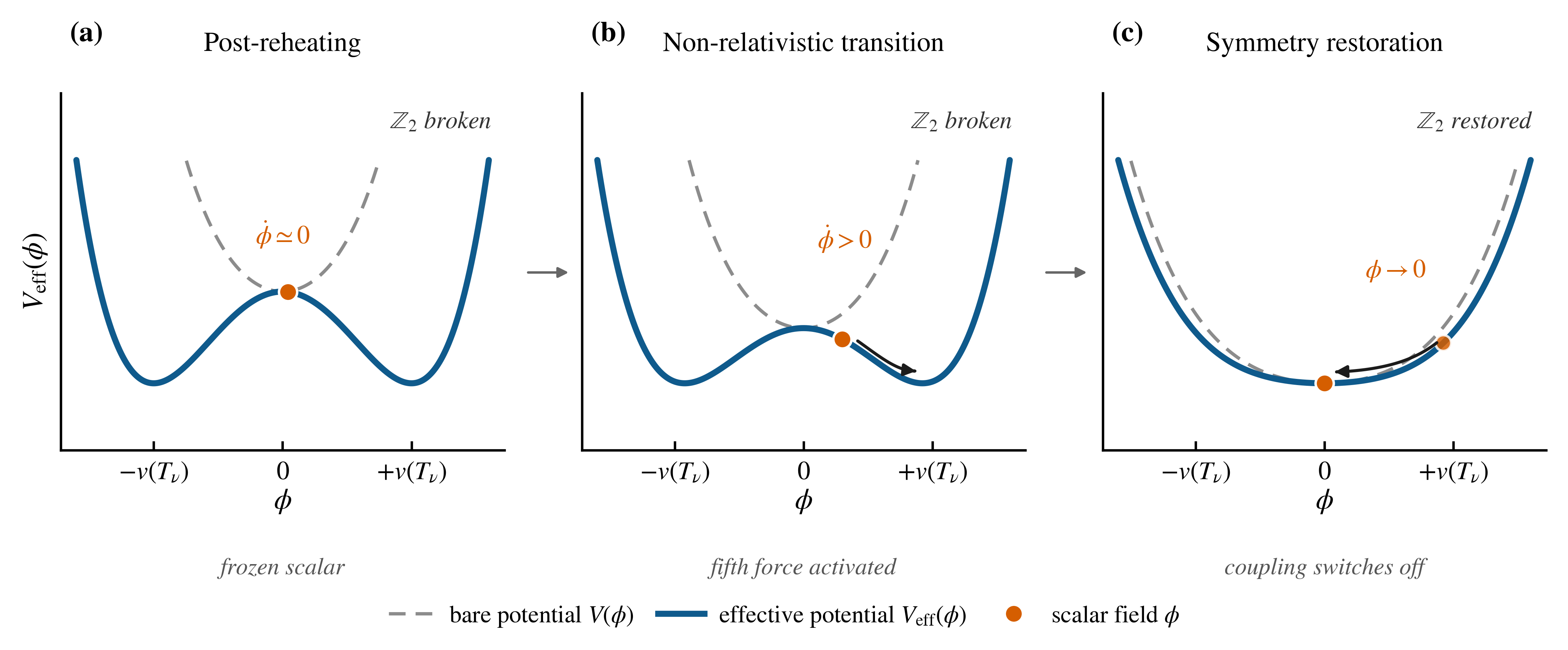}
    \caption{Behavior of the scalar field coupled to neutrinos in the inverse phase transition scenario.}
    \label{fig:reverse_scenario}
\end{figure*}

The source term relating to the interaction in the Klein-Gordon equation \eqref{eq:KG_bis} changes sign
\begin{align}
\label{eq:Klein_Gordon_anti}
\ddot{\phi}+2 \mathcal{H} \dot{\phi}+a^2 V_{,\phi}
&=-a^2 \frac{d \ln m_\nu}{d\phi}\left(\rho_\nu - 3p_\nu\right)\,,\\
&=+a^2 \frac{\phi}{M^2}\left(\rho_\nu - 3p_\nu\right)\,.
\end{align}
The effective potential of the scalar field now reads
\begin{align}
\label{eq:effective_potential_anti}
    V_{\mathrm{eff}} = 
    \frac{1}{2}\mu^2 \phi^2
    + \frac{\lambda}{4}\phi^4 +\left(\rho_\nu - 3p_\nu\right) \ln A(\phi)+\rho_\Lambda\,,
\end{align}
and its derivative becomes,
\begin{equation}\label{eq:dVeff_anti}
    V^\prime_{\mathrm{eff}}(\phi)=\left( -\frac{\rho_\nu-3p_\nu}{M^2}+\mu^2 \right)\phi+\lambda\phi^3\,.
\end{equation}
The effective mass at $\phi = 0$ in the inverse phase transition scenario is
\begin{equation}
    \label{eq:meff_anti}
    m^2_{\mathrm{eff}} \equiv V''_{\mathrm{eff}}\big|_{\phi=0}
    = \mu^2 + \frac{T_\nu}{M^2}\,,
\end{equation}
which has the opposite sign structure compared to the standard symmetron
case in Eq.~\eqref{eq:meff_standard}. 
Figure~\ref{fig:reverse_scenario} illustrates the effective potential evolution in a typical scenario where the field is now initialized at $\phi(z_\mathrm{ini})\approx0$ before reheating, in Panel (a). Symmetry breaking ($m^2_{\mathrm{eff}} < 0$) occurs when $-T_\nu>\rho_\star$ while the field motion is initially overdamped by the expansion. Once unfrozen the field rolls towards one of the vacuum expectation values, in Panel (b). At late time, as the Universe expands and
$\rho_\nu < \rho_\star$, symmetry is restored, in Panel (c).

We notice that since the effective potential in Eq.~\eqref{eq:effective_potential_anti} comes with a logarithm, it looks like the Taylor expansion source that we considered before, at the level of the field equations of the standard symmetron. The evolving minimum of this potential is
\begin{equation}
\label{eq:evolving_minimum}
    v_\pm = \pm v_0 \sqrt{\mathrm{max}\left\{-\frac{T_\nu}{\rho_\star}-1,
    \;0\right\}}\,,
\end{equation}
where $v_0=\mu/\sqrt{\lambda}$ and $\rho_\star= \mu^2 M^2$ is the critical energy density of the phase transition, as before in Eq.~\eqref{eq:v_0}. In the nonrelativistic limit, ${T_\nu\simeq-\rho_\nu}$, so that $|v|$ grows monotonically with $\rho_\nu$.

\begin{figure*}[t]
    \centering
\includegraphics[width=0.49\linewidth]{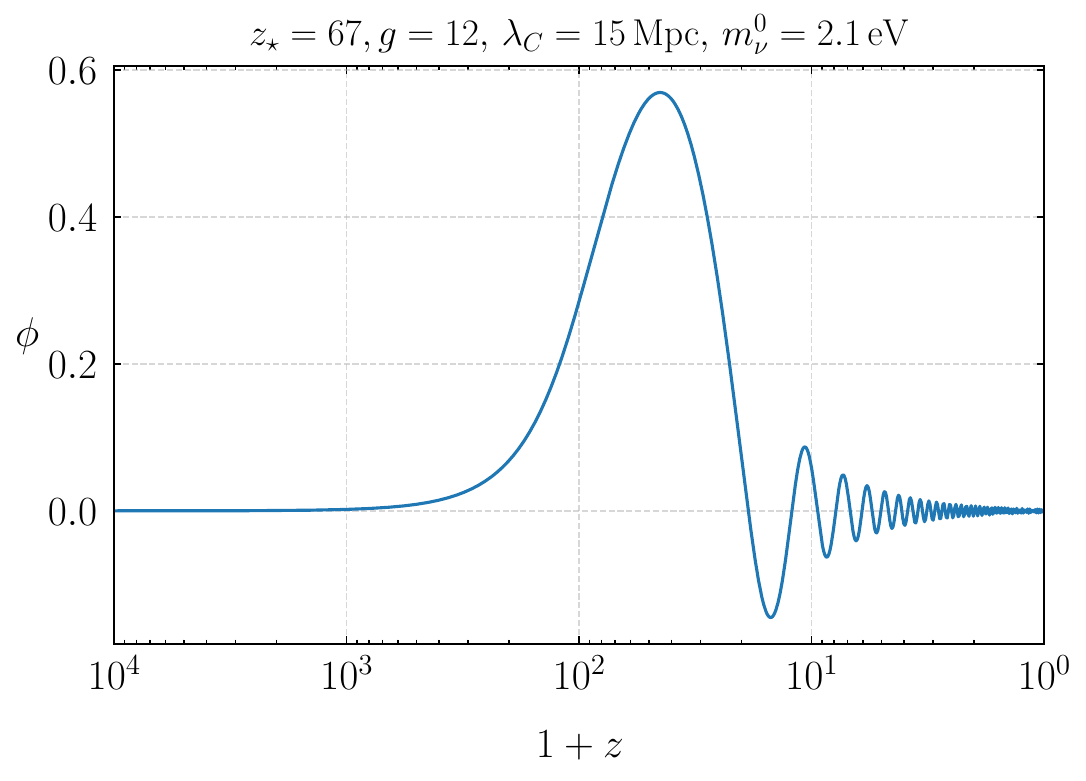}
\includegraphics[width=0.49\linewidth]{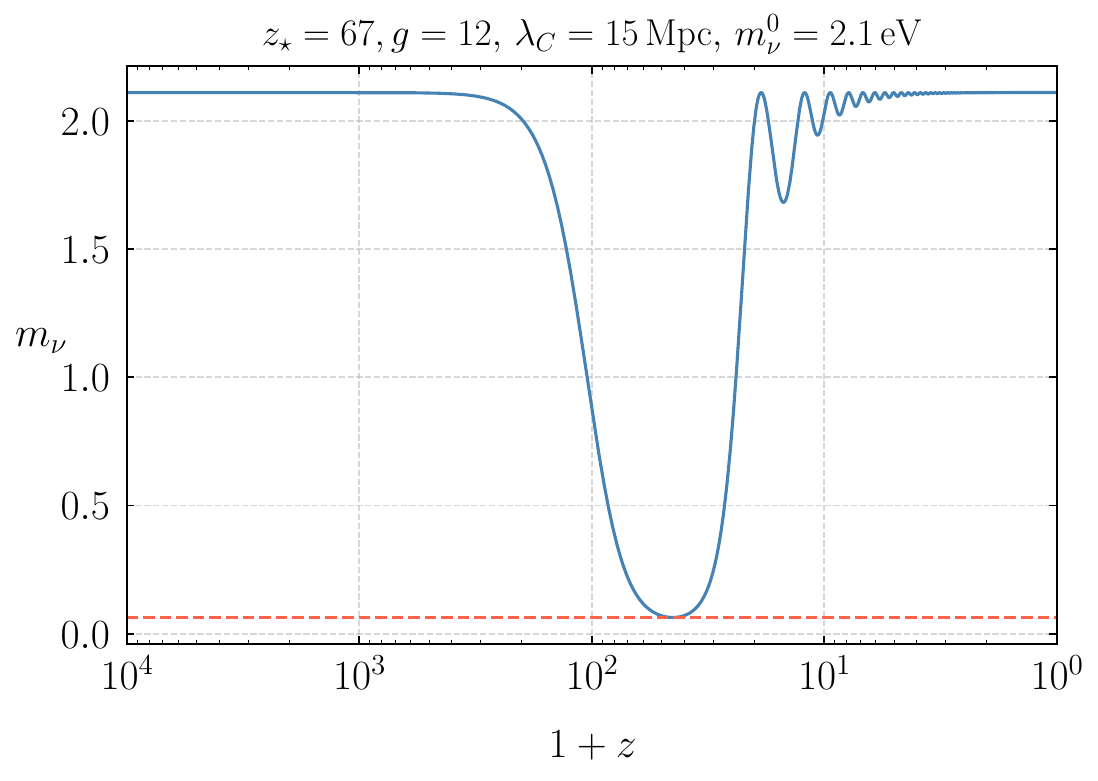}
\caption{Left: Evolution of the scalar field in the inverse phase transition model. Right: Evolution of the neutrino mass in the inverse phase transition model. The horizontal dashed red line corresponds to $m_\nu=0.06\,\mathrm{eV}.$}
\label{fig:phi_anti}
\end{figure*}

As in the left panel of Fig.~\ref{fig:phi_anti}, an unfrozen scalar field acquires a nonzero vacuum expectation value $v_\pm$ for nonrelativistic neutrinos with $\rho_\nu>\rho_\star$. This activates the scalar-neutrino coupling and allows the neutrino mass to vary. During this phase, the mass decreases as the field moves away from $\phi=0$. In the example illustrated in the right panel of Fig.~\ref{fig:phi_anti}, the neutrino mass decreases from the bare value $\hat m_\nu=2.1\,\mathrm{eV}$ to $m_\nu=0.06\,\mathrm{eV}$. The minimum neutrino mass depends on the ratio \(\rho_{\nu,\max}/\rho_\star\), the parameter ratio \(v_0/M\), and the extent to which the scalar field follows the evolving minimum of the effective potential.

As the Universe expands, $\rho_\nu$ eventually drops below the critical value ${\rho_\star=\mu^2M^2}$. The symmetry is then restored, the effective minimum returns to ${\phi=0}$, and the scalar-neutrino coupling dynamically switches off. As a result, the neutrino mass increases back to its bare value, ${m_\nu^0=\hat m_\nu=2.1\,\mathrm{eV}}$, provided the field has returned to $\phi=0$ today.

\begin{figure}[t]
    \centering
\includegraphics[width=0.7\linewidth]{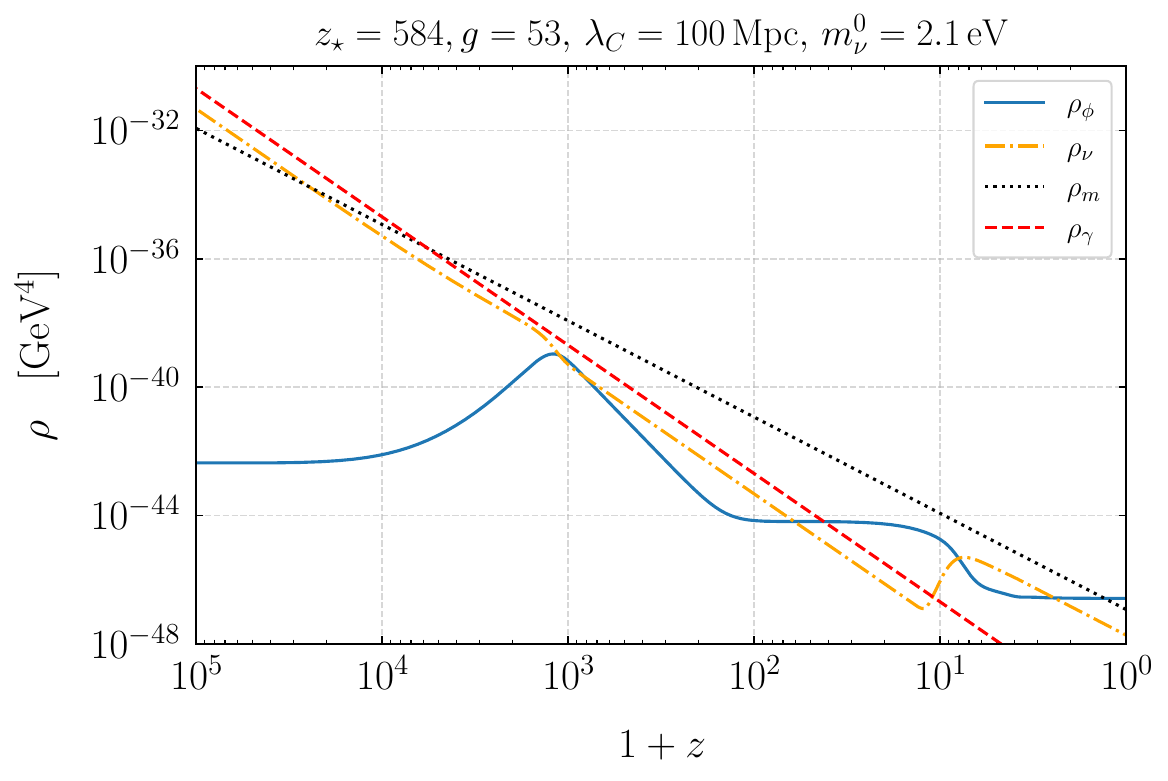}
    \caption{Evolution of the energy densities for photons ($\rho_\gamma$), matter ($\rho_m$), neutrinos ($\rho_\nu$), and the scalar field ($\rho_\phi$) in the inverse phase transition scenario.}
\label{fig:sym_densities_anti}
\end{figure}

In the inverse phase transition scenario the sign of the interaction term in the scalar continuity equation is opposite to that in Eq.~\eqref{eq:continuity_scalar}:
\begin{equation}
\label{eq:continuity_scalar_inverse}
    \dot{\rho}_\phi+3\mathcal{H}\left(\rho_\phi+p_\phi\right)=+\frac{\phi\dot{\phi}}{M^2}\left(\rho_\nu-3p_\nu\right)\,,
\end{equation}
and the effective scalar equation of state, $w_\phi^\mathrm{eff}$, fulfills
\begin{equation}
    w_\phi^\mathrm{eff}=w_\phi-\frac{\phi\dot{\phi}}{3\mathcal{H}M^2}\frac{\rho_\nu-3p_\nu}{\rho_\phi}\,.
\end{equation}
Figure~\ref{fig:sym_densities_anti} shows that the scalar energy density can temporarily  increase -- a sign of a phantom effective equation of state $w_\phi^\mathrm{eff}<-1$ -- even when ${w_\phi>-1}$, if the coupling term is positive. The corresponding net energy transfer from neutrinos to the scalar field can lead to $w_\phi^{\rm{eff}}<-1$. In the example shown, the field energy density peaks around recombination, \(z\sim 10^3\), before the scalar component becomes largely subdominant again. Such a localized injection increases the pre-recombination expansion rate and reduces the comoving sound horizon. Since cosmic microwave background observations tightly constrain the angular sound scale, a smaller sound horizon can be compensated for with a reduced angular-diameter distance to last scattering. This, in turn, allows for a larger inferred value of the Hubble parameter \(H_0\). Thus, the neutrino-scalar interactions in this model may provide a mechanism to alleviate the Hubble tension \cite{DiValentino:2025sru}, analogously to neutrino-assisted early dark energy scenarios \cite{Sakstein:2019fmf,deSouza:2023sqp,CarrilloGonzalez:2023lma}.

\subsection{Strength of the fifth force}
In the inverse phase transition model, the relative strength of the fifth force does not saturate to the value in Eq.~\eqref{eq:g} due to the evolving potential minimum $v_\pm$ given by Eq.~\eqref{eq:evolving_minimum}. To find the respective fifth force, we consider the Jordan frame Bardeen potential $\tilde g_{tt}=-e^{2\tilde \Psi}$. Since $\tilde g_{tt}=A^2g_{tt}$, this means that
\begin{equation}
    \tilde \Psi = \Psi+\ln A\,.
\end{equation}

It is assumed for now that the metric allows a weak-field expansion in the Jordan frame. For the nonrelativistic, weak-field geodesic equation for a particle in Jordan frame (where particles move on geodesics)
\begin{equation} \label{eq:fifthNewtonian}
    \ddot x = \partial_r \tilde \Psi = \partial_r \Psi + \frac{\partial _r A}{A}\,,
\end{equation}
so that we again produce similar contributions in the exponential and Taylor models, owing to the logarithm. A fully relativistic and non-perturbative calculation gives the same result for the Einstein frame, nonrelativistic $v\ll 1$, weak-field expanded limit $\Psi=\Phi, \, B_i\sim h_{ij}\sim 0$, without assumptions about weak fields in the Jordan frame, which is shown in a work in preparation. For our applications, where $A\nsim 1$,  
weak fields in the Einstein frame may be strong in the Jordan frame. Equation~\eqref{eq:fifthNewtonian} translates to a relative fifth force strength 
\begin{equation}
    \frac{F_5}{F_N} = - \frac{v\;\partial_r\phi}{M^2 \;\partial_r\Psi}\,, \label{eq:forceeq1}
\end{equation}
where $F_5=\partial_r\log A$ is the fifth force mediated by the scalar field and $F_N=\partial_r \Psi$ is the gravitational force. From the quasistatic limit of the scalar field equation, when perturbing the density, 
\begin{equation}
    \Delta \phi = a^2 V_\phi^{\mathrm{eff}} (v_+;\bar\rho+\delta\rho) = a^2\delta\rho\; \partial_\rho V_{\phi}^{\mathrm{eff}}(v_+;\bar\rho)\,, 
\end{equation}
where $V_\phi^{\mathrm{eff}}(v_+;\bar\rho) = 0$ since it is a minimum. This yields
\begin{equation}
    \Delta\phi = -a^2 \delta\rho \frac{v_+}{M^2}\,,
\end{equation}
which when restoring the Planck mass, gives us a quantity a factor $2\, v_+/M^2$ larger than the source of the Newtonian Poisson equation (and of opposite sign). Therefore, when integrating,  $\partial_r\phi\sim -(2 v_+/M^2)\,\partial_r\Psi$. Inserting this into Eq.~\eqref{eq:forceeq1}, we see 
\begin{equation}
    \frac{F_5}{F_N} =  2\left(\frac{v_+}{M^2}\right)^2=2g^2(\rho_\nu)\,,
\end{equation}
where we define 
\begin{equation}
    g(\rho_\nu)\equiv\frac{v_+\,M_{\rm Pl}}{M^2}\,,
\end{equation}
with $v_+$ given in Eq.~\eqref{eq:evolving_minimum}. Notably, the negative sign disappears, confirming that the fifth force remains attractive. A repulsive scalar force would require us to change the sign of the Laplacian term, which corresponds to a ghost.

\subsection{Impact on perturbations}

In the inverse symmetry-breaking pattern, the coupling is only active during a limited cosmological epoch. This allows the neutrino mass to vary and potentially influence the background expansion and structure formation. However, at later times, the scalar field  evolves toward a regime where the effective coupling is suppressed, shutting down the fifth force. This is expected to prevent excessive growth of neutrino perturbations and avoid late-time instabilities associated with a persistently strong scalar-mediated interaction. Thus, neutrino mass variation could occur during a controlled period in cosmic history while the fifth force naturally disappears at late times, restoring standard gravitational clustering.

To show that this scenario prevents late-time instabilities encountered in the original setup, we implement the model in the CLASS solver by adapting the code. In particular, the relevant equations are modified with the effective mass in Eq.~\eqref{eq:mass_anti}, and its derivatives that concern the coupling terms,
\begin{align}
    \frac{d\ln m_\nu}{d\phi}&=-\frac{\phi}{M^2}\,,\\
    \frac{d^2\ln m_\nu}{d\phi^2}&=-\frac{1}{M^2}\,.
\end{align}
Note that the minus sign is the only difference with the standard symmetron model.

\begin{figure}[t]
    \centering
\includegraphics[width=0.7\linewidth]{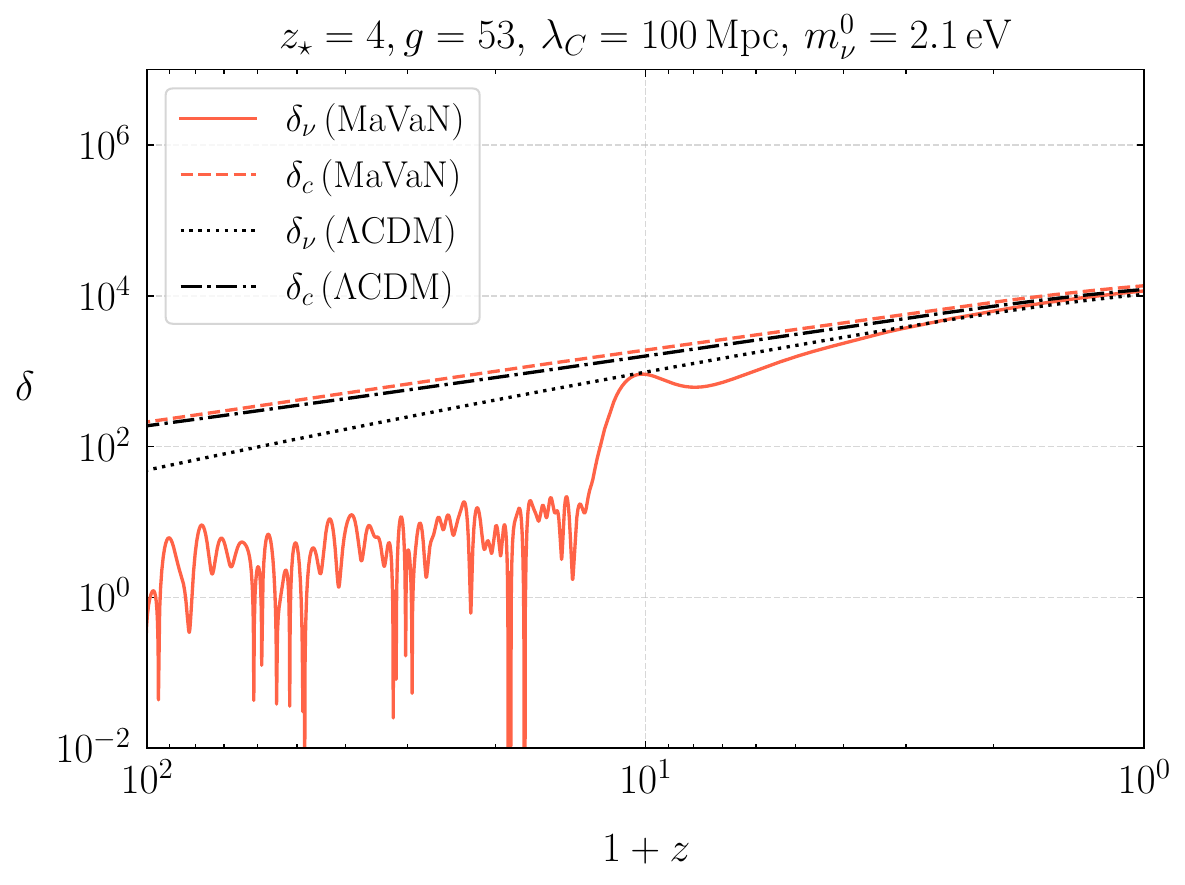}
    \caption{Evolution of the neutrino $(\delta_\nu)$ and cold dark matter $(\delta_c)$ density contrast for the MaVaN inverse phase transition model and $\Lambda\mathrm{CDM}$ (with neutrinos of mass $m_\nu^0=2.1\,\mathrm{eV}$), at scale $k=0.1\,\mathrm{Mpc}^{-1}$, normalized to the primordial curvature amplitude \(\mathcal{R}=1\).}
\label{fig:density_contrast_anti}
\end{figure}

The mass is growing at late times when the interaction is switching off. In this example, after the neutrinos become nonrelativistic, the effective mass decreases down to $m{_\nu\simeq3.5\times10^{-7}\,\mathrm{eV}}$ during the broken phase. Once the symmetry is restored, the field returns to ${\phi=0}$, the coupling switches off, and the mass increases back to its present-day value, $m_\nu^0=2.1\,\mathrm{eV}$. But since the fifth force disappears, the growth of neutrino perturbations are driven by gravity alone and the model avoids the instability associated with the large scalar-mediated enhancement of neutrino clustering. This is illustrated in Fig.~\ref{fig:density_contrast_anti} for parameter values similar to those that produce the instabilities in the standard symmetron model.

\section{Discussion}\label{sec:conclusions}

Our goal was to build a MaVaN scenario that could avoid the well-known instabilities associated with large neutrino masses and strong scalar couplings. In standard realizations, large neutrino densities and strong fifth forces usually trigger excessive growing modes for both neutrino and dark matter perturbations. They eventually exhibit unphysically large contrasts at late times on small linear scales.

Our approach is based on a symmetry-driven screening mechanism. However, within the standard symmetron framework, we found that linear instabilities persist at the background level for early phase transitions. Non-linear symmetron screening may reduce or halt instability when neutrino overdensities exceed a critical screening threshold. However, determining whether this is sufficient to make the model observationally viable requires analyzing the non-linear regime. We leave this question to a future project.

To improve the linear behavior of the model, we introduced an alternative scenario, which we refer to as an inverse phase transition. Concretely, we proposed a conformal coupling function in the form of a negative exponential, while the quadratic term in the bare scalar potential is positive (opposite to the standard symmetron case). The nonrelativistic transition of the neutrinos triggers symmetry breaking and activates the coupling associated with the fifth force. As the Universe expands and the neutrino density subsequently drops below a critical threshold, symmetry is restored and the coupling is suppressed. This late-time deactivation of the fifth force prevents the excessive growth of neutrino perturbations and avoids late-time instabilities in the linear regime.

Interestingly, the coupling can generate a non-negligible early dark energy component around recombination. This component has the potential to alleviate the Hubble tension by reducing the sound horizon at decoupling and allowing for larger inferred values of $H_0$ from high redshift probes. A thorough exploration of the parameter space through Bayesian inference would be needed to assess whether current cosmic microwave background data favor such scenarios. We leave this investigation for future work.

Finally, it would be worth studying the behavior of the  inverse phase transition model in the non-linear regime, since in regions of sufficiently high density, symmetry breaking may reoccur or persist. 
This could potentially trigger a runaway effect, whereby enhanced clustering strengthens the fifth force, which in turn further amplifies gravitational collapse. Such dynamics could have significant implications for neutrino clustering, structure formation, and the overall stability of the model, which requires further analysis.

\clearpage


\acknowledgments
The authors thank Julian Adamek for his comments and Farbod Hassani for his help clarifying the CLASS unit conventions. This work is supported by Fundação para a Ciência e a Tecnologia (FCT) through national funds under the R\&D Unit grant UID/04434/2025 (DOI 10.54499/UID/04434/2025). VdF acknowledges support from FCT through grant 2022.14431.BD. ØC acknowledges support from the European Union and the Czech Ministry of Education, Youth and Sports (Project No. FORTE – CZ.02.01.01/00/22\_008/0004632). DFM thanks the Research Council of Norway for their support and the resources provided by UNINETT Sigma2-the National Infrastructure for High-Performance Computing and Data Storage in Norway.

\bibliographystyle{JHEP}
\bibliography{bib,inspire}

\end{document}